\documentclass[14pt, onecolumn]{aa}
\usepackage{graphicx}
\usepackage{txfonts}
\usepackage{natbib}
\bibpunct{(}{)}{;}{a}{}{,}
\begin{document}

\title{A bright coronal downflow seen in multi-wavelength observations: 
evidence of a bifurcating flux-rope?}
   

   \author{D. Tripathi\inst{1}, S.~K. Solanki\inst{2},
   H.~E. Mason\inst{1} and D.~F. Webb\inst{3}}

   \offprints{D.Tripathi@damtp.cam.ac.uk}

   \institute{Department of Applied Mathematics and Theoretical
     Physics, Wilberforce Road, Cambridge, CB3 0WA, UK\\
    \email{[D.Tripathi; H.E.Mason]@damtp.cam.ac.uk}
    \and Max-Planck-Institut f\"ur Sonnensystemforschung, 37191 
    Katlenburg-Linda, Germany\\
    \email{solanki@mps.mpg.de}
    \and Institute for Scientific Research, Boston College, Chestnut
     Hill, Messachusetts, USA\\
    \email{David.Webb.ctr@hanscom.af.mil}}

   \date{Accepted for publication in Astronomy and Astrophysics}

\abstract{}{To study the origin and characteristics of a bright
coronal downflow seen after a coronal mass ejection associated with
erupting prominences on 5~March 2000.}{This study extends that of
Tripathi et al. (2006b) based on the Extreme-ultraviolet Imaging
Telescope (EIT), the Soft X-ray Telescope (SXT) and the Large Angle
Spectrometric Coronagraph (LASCO) observations. We combined those
results with an analysis of the observations taken by the H${\alpha}$
and the Mk4 coronagraphs at the Mauna Loa Solar Observatory
(MLSO). The combined data-set spans a broad range of temperature as
well as continuous observations from the solar surface out to 30
R$_{\sun}$.}  {The downflow started at around 1.6R$_{\sun}$ and
contained both hot and cold gas. The downflow was observed in the
H${\alpha}$ and the Mk4 coronagraphs as well as the EIT and the SXT
and was approximately co-spatial and co-temporal providing evidence of
multi-thermal plasma. The H${\alpha}$ and Mk4 images show cusp-shaped
structures close to the location where the downflow started. Mk4
observations reveal that the speed of the downflow in the early phase
was substantially higher than the free-fall speed, implying a strong
downward acceleration near the height at which the downflow
started.}{The origin of the downflow was likely to have been magnetic
reconnection taking place inside the erupting flux rope that led to
its bifurcation.}


\titlerunning{Bright Coronal Downflow: Bifurcating flux-rope?} 

\authorrunning{Durgesh Tripathi et al. }

\maketitle


\section{Introduction} 

Observations of coronal downflows in X-rays \citep{mac99, mac00}, EUV
radiation \citep{davina2,davina1,asai04_down} and in white-light
\citep{wang_inflow, sheeley_inflow} after Coronal Mass Ejections
(CMEs) were first detected based on observations made by the Soft
X-ray Telescope \citep[SXT;][]{sxt} aboard Yohkoh, the Transition
Region and Coronal Explorer \cite[TRACE;][]{trace} and the Solar
Ultraviolet Measurements of Emitted Radiation \citep[SUMER;][]{sumer}
spectrometer and the Large Angle Spectrometer Coronagraph
\citep[LASCO;][]{lasco} aboard the Solar and Heliospheric Observatory
\citep[SoHO;][]{soho}, respectively. These downflows appeared to be
{\it dark} and were interpreted as plasma voids with high temperature
and low density as a consequence of magnetic reconnection following
the CME eruptions.

\begin{figure}
\centering
\includegraphics[width=0.45\textwidth]{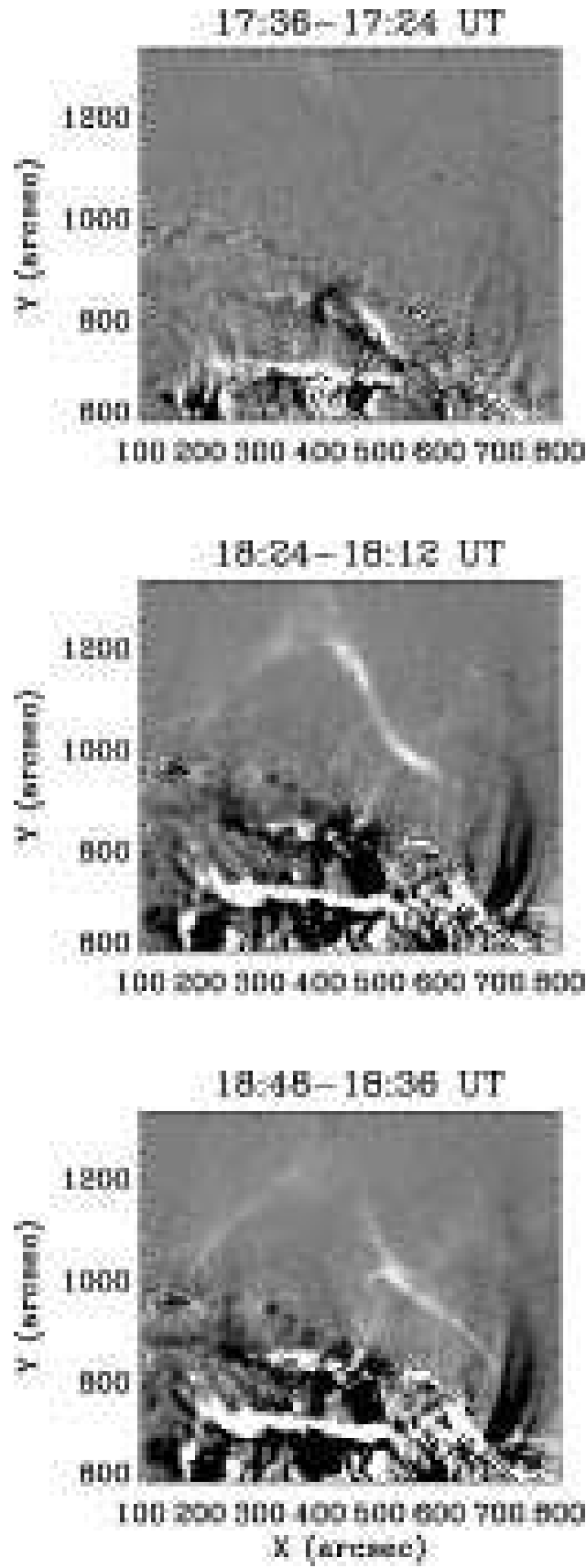}
\caption{Running difference images taken by EIT at 195~{\AA} on
5~March 2000 showing the {\it bright} coronal downflow. In the top
image it is just visible at the top of the panel at x$\approx$400
arcsec. \label{eit_downflow}}
\end{figure}

Recently, \cite{trip_inflow06}, referred to as ``Paper~I''
hereinafter, reported a {\it bright} coronal downflow after a CME
event, which occurred on 5~March 2000. This was the first {\it bright}
coronal downflow following a CME eruption observed at EUV wavelengths
by the Extreme-ultraviolet Imaging Telescope \citep[EIT;][]{eit,
mos_eit} also aboard the SoHO. Based on EIT 304 {\AA}
\cite{groof_a} presented bright coronal downflows in coronal loop
without any associated eruption. The origin of these downflows was
explained by numerical simulations of "catastrophic cooling" in a
coronal loop which is heated predominantly at its footpoints
\citep{groof_b}.

In contrast to the dark downflows following eruption, the bright
downflow indicates a flow of heated plasma, thus providing more direct
evidence of magnetic reconnection during the eruption of CMEs. The
corresponding CME was associated with three erupting
prominences. Based on the analysis of the images obtained by EIT at
195~{\AA} and by SXT, it was speculated that the downflow could indeed
be a consequence of magnetic reconnection, taking place somewhere
outside the field-of-view (FOV) of EIT, but behind the occulter of the
LASCO/C2. Fig.~\ref{eit_downflow} displays the running difference
images of the downflow taken by the EIT at 195~{\AA}. The fact that
the downflow started in the gap between the area covered by two
instruments leads to significant uncertainties in the
interpretation. We have therefore searched other databases with the
aim of finding data covering this uncharted region. This event was
fortuitously also recorded by the H${\alpha}$ coronagraph, the Mk4
coronameter and the He 10830~{\AA} telescope at Mauna Loa Solar
Observatory
(MLSO\footnote{http://mlso.hao.ucar.edu/cgi-bin/mlso\_homepage.cgi})
in Hawaii. These telescopes nicely fill the gap between the EIT and
the LASCO C2. While MLSO did not catch the eruption phase, it did
provide observations of the downflow.

Observations of downflows in the majority of erupting prominences were
reported by \cite{holly00, holly01} based on the data taken with HAO's
(High Altitude Observatory) MLSO instruments. According to
\cite{holly00, holly01}, in the process of eruption the prominences
break into two parts involving the formation of an X-type neutral line
and magnetic reconnection. The separation seemed to take place at a
height ranging from 1.20 to 1.35 R$_{\sun}$. However, there was no
direct evidence for magnetic reconnection and the formation of
an X-type neutral line. Moreover, no attempts were made to compare the
H${\alpha}$ observations with data taken by other instruments such as
SoHO/EIT or Yohkoh/SXT.

Based on the observations made by the H${\alpha}$ coronagraph at MLSO,
\cite{holly01} discussed that, for the flux-rope type of topology,
reconnection could take place either below or within the flux rope
during the eruption. Generally, in 'standard 2D models' it is
considered that the magnetic field lines overlying the flux-rope
reconnect at the current sheet in the wake of expulsion of the
flux-rope \citep[e.g.,][\& references therein]{lin_forbes00}. In this
scenario the total expulsion of the flux rope occurs and the flux rope
propagates into the interplanetary medium along with the corresponding
CME. Note that in this scenario the magnetic field lines forming
the flux rope do not take part in reconnection. On the other hand,
the reconnection could also take place 'internally', i.e., within the
flux rope  where field lines forming the flux rope take part in
reconnection \citep{manch_04, sarah_part07, sarah_part06}, leading to
the bifurcation of the flux rope during eruption.

\begin{figure*}
\centering
\includegraphics[width=0.75\textwidth]{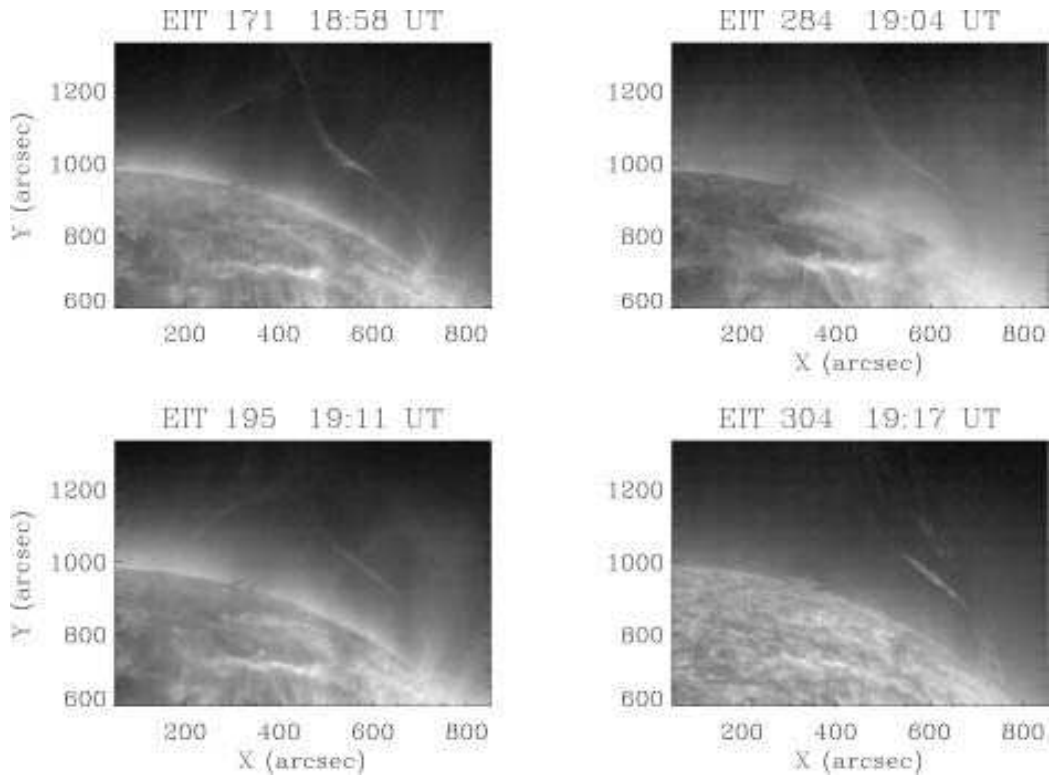}
\caption{Absolute intensity images recorded by the EIT in its 171~{\AA}
(top left), 284~{\AA} (top right), 195~{\AA} (bottom left) and
304~{\AA} (bottom right) channels. Note that the images are displayed
on a logarithmic scale. \label{eit_four}}
\end{figure*}

The observations and hypothesis made by \cite{holly00, holly01} were
later studied by \cite{sarah_part07, sarah_part06} based on a 3D MHD
simulation. In these simulations the complete evolution of a flux rope
was studied from the solar surface out to 6 R$_{\sun}$. Based on the
results of their simulation, \cite{sarah_part07, sarah_part06} found
that the emergence of a flux rope with enough twist causes it to erupt
due to loss of equilibrium. The flux rope undergoes a kink
instability which leads to the formation of a vertical current sheet
inside the flux rope. The formation of a current sheet within an
unstable flux rope has also been demonstrated by \cite{birn}. After
multiple reconnections occurring inside the flux rope at the current
sheet, formed during the eruption, the flux-rope breaks in two. One
part of the flux rope escapes as the core of a corresponding CME and
the other falls back towards the Sun's surface. However, in the
simulation presented by \cite{sarah_part07, sarah_part06}, a special
kind of magnetic field geometry was used - {\it a Bald Patch
Separatrix Surface} (BPSS) - where the flux-rope intersects the

In this paper we present observations of a bright coronal downflow
which may be an example of such a bifurcating flux rope in the course
of a CME eruption. Here we investigate the observation of the downflow
taken in multiple wavelengths, such as H${\alpha}$, white-light
K-corona (Mk4), and EUV. In the next section, we present the
observations we used, followed by their analysis and results in
section 3. In section 4, we provide measurements performed on the
data. We provide a summary of the results and discussion in section 5.

\section{Observations}
\begin{figure*}
\centering
\includegraphics[width=0.75\textwidth]{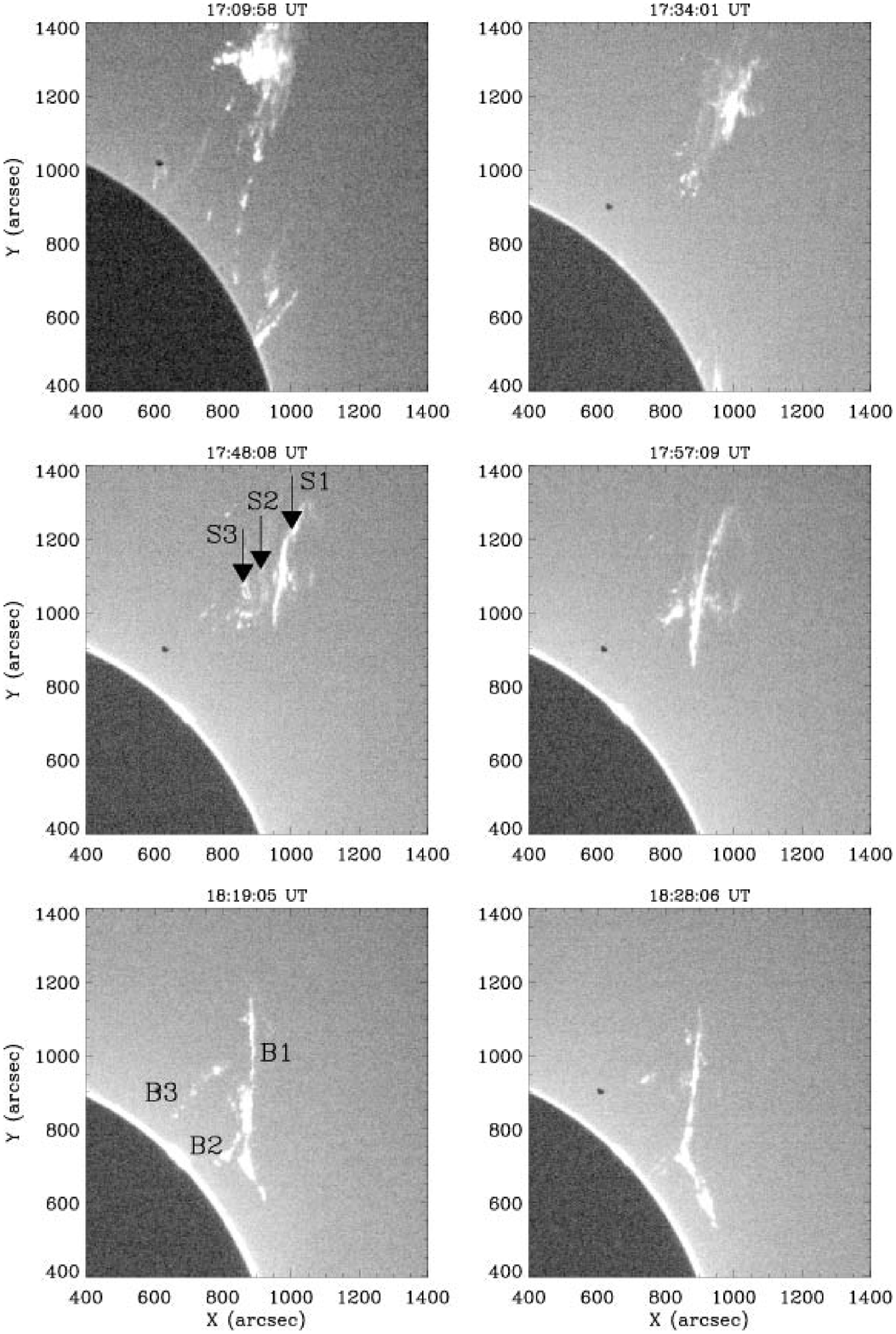}
\caption{Sequence of images taken by the $H{\alpha}$ coronagraph of
the Advanced Corona Observing System at Mauna Loa Solar Observatory
showing the complete morphology of the downflow. The images are
displayed on a logarithmic scaling of intensity. Note that in this
figure only some selected images are shown. For the complete evolution
of the downflow see the movie 'halpha.mov' (online
only).}\label{halpha}
\end{figure*}
\begin{figure}
\centering
\includegraphics[width=0.4\textwidth]{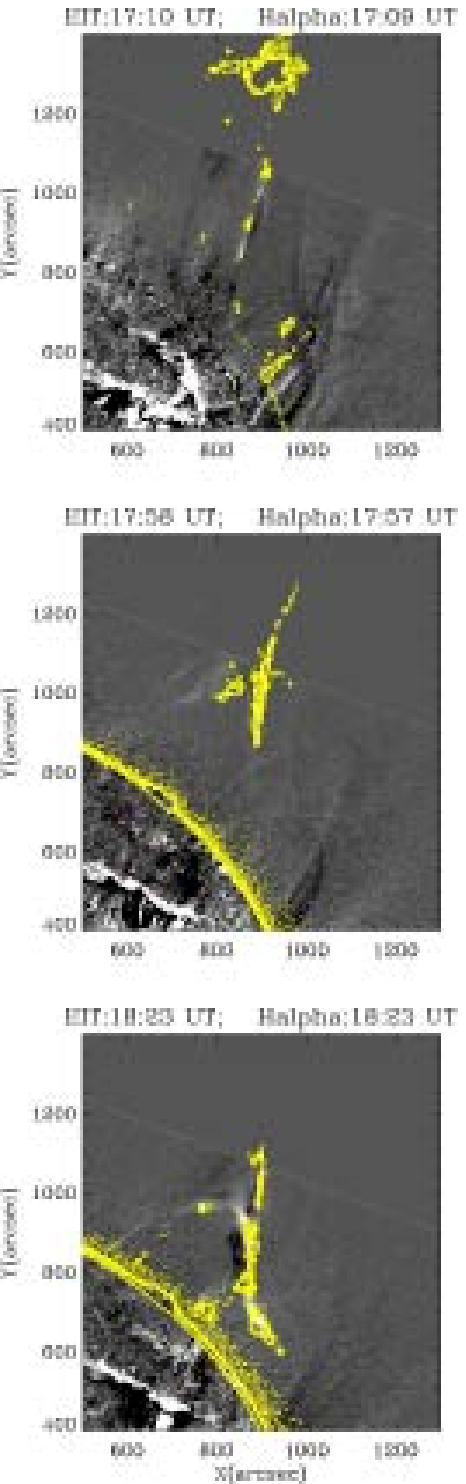}
\caption{Co-aligned over-plotted EIT (running difference; in black \&
white) and $H{\alpha}$ images (yellow contours).}\label{eit_halpha}
\end{figure}

Three erupting prominences (two large and one small - barely
discernible) followed by a coronal downflow were observed on
05~March 2000 by the EIT. The EIT provides
observations of the Sun at 195~{\AA} with a regular cadence of about
12~mins and one image every 6 hours at 171~{\AA}, 284~{\AA} and
304~{\AA}. The 195~{\AA} passband of EIT is dominated by an
\ion{Fe}{xii} line formed at 1.5~MK, but also contains an
\ion{Fe}{xxiv} line at 192~{\AA} formed at around 20~MK, which is
usually much weaker in the quiet Sun region but highly significant in
flaring regions \citep{trip_trace}. The images obtained by the EIT at
171~{\AA}, 284~{\AA} and 304~{\AA} wavelengths are dominated by lines
\ion{Fe}{ix}{/x} (1.0~MK), \ion{Fe}{xv} (1.8~MK) and \ion{He}{ii}
(0.05~MK) respectively.

Fortuitously this downflow was also observed by the Advanced Corona
Observing System
(ACOS)\footnote{http://mlso.hao.ucar.edu/cgi-bin/mlso\_data.cgi?2000\&ACOS}
composed of the Polarimeter for Inner Coronal Studies (PICS)
H${\alpha}$ (6563~{\AA}) coronagraph, the Mk4 K-coronameter, which
observes the white-light K-corona, and the Chromospheric Helium \rm{I}
Imaging Photometer (CHIP) \ion{He}{i} (10830~{\AA}) instrument. In
this paper we concentrate on observations taken by the H${\alpha}$
coronagraph and the Mk4 coronameter. Since the downflow was not
clearly seen in the CHIP data, we decided not to use it.

\begin{figure}
\centering
\includegraphics[width=0.6\textwidth]{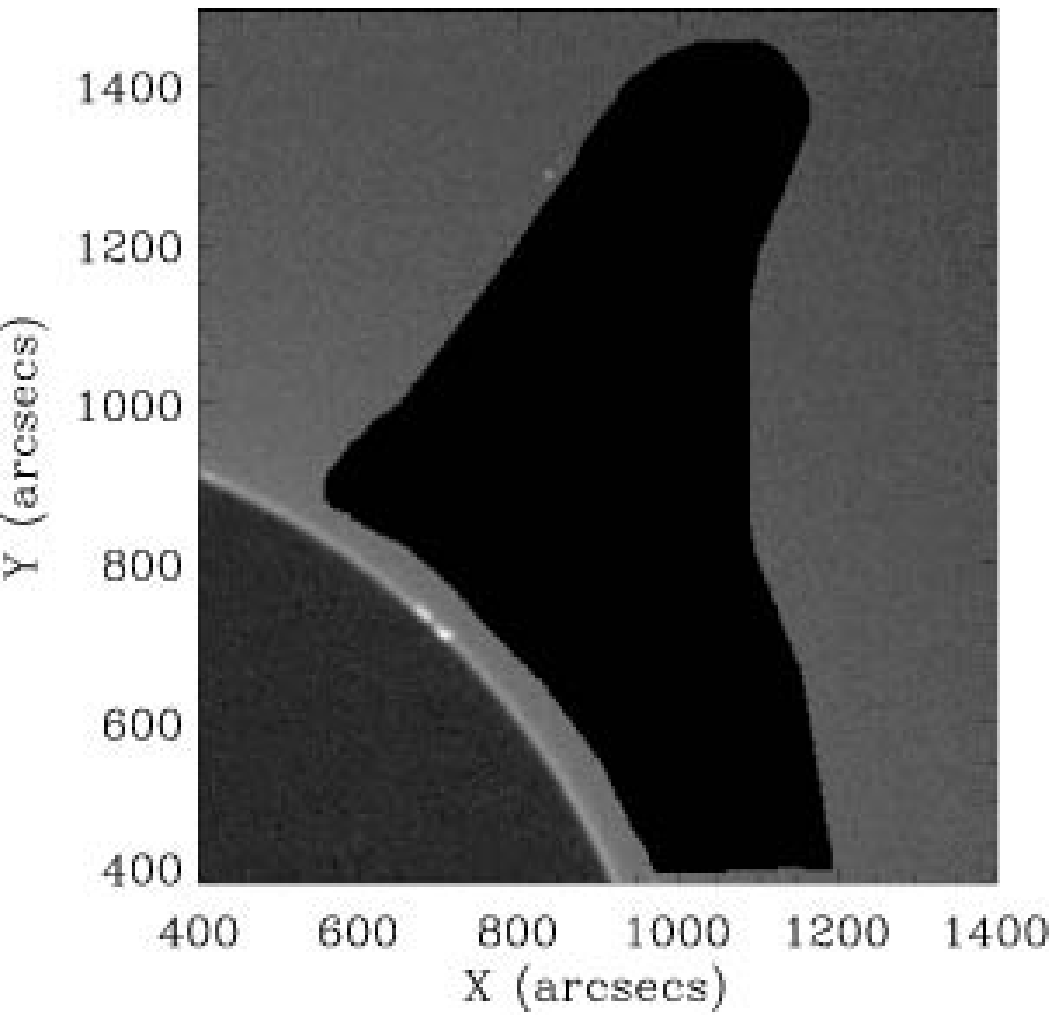}
\includegraphics[width=0.5\textwidth]{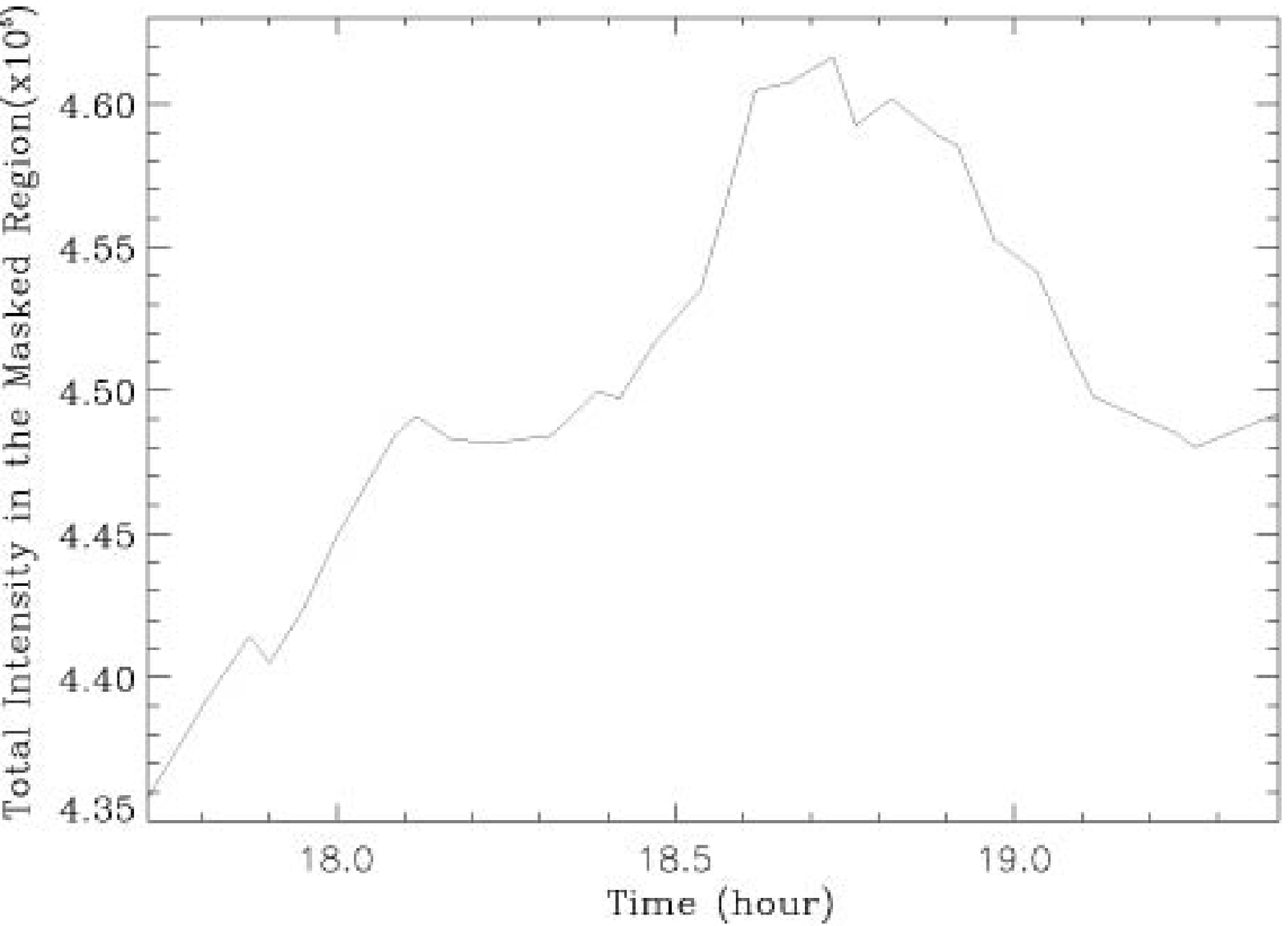}
\caption{Top panel: Masked H${\alpha}$ image. Bottom panel: variation
of total intensity in the masked region with time. \label{mask}}
\end{figure}

The ACOS instruments are operated at MLSO in Hawaii by the HAO. They
operate every day from about 17:00 until 22:00~UT (weather permitting)
producing about 100 images for each instrument with a cadence of about
3 mins.  The field of view (FOV) ranges from 1.01 to 1.83 R$_{\sun}$
from the solar center for the H${\alpha}$ coronagraph and from 1.12 to
2.79 R$_{\sun}$ for Mk4. The pixel size of the images taken by the
H${\alpha}$ coronagraph and Mk4 coronameter is about 2.9 arcsec. For
white-light observations of the corona, the Mk4 coronameter records
the polarization brightness data as well as white-light vignetted
data, where a hypothetical density function is subtracted from the
actual data in order to enhance the contrast in the images. Since we
are interested in a morphological study, we consider the Mk4
white-light vignetted data in this paper. The Mk4 data provided in the
archive are essentially fully processed. For technical details
concerning data processing and the instruments see \cite{elmore}.

\section{Data analysis and results}
About 90 minutes (at 17:36~UT) after the onset of the eruption, a
bright feature appeared at the edge of the FOV of the EIT. After a
while, this {\it bright} feature propagated towards the Sun's surface
and is referred to as the {\it bright coronal
downflow}. Fig.~\ref{eit_downflow} displays a series of running
difference images taken by the EIT at 195~{\AA}, revealing the
morphological evolution of the downflow in time. Since the EIT
observations are described in detail in Paper I, we restrict ourselves
to saying that the three images show from top to bottom the initial
appearance of the downflowing material at the edge of the EIT FOV, an
intermediate phase and a late phase of the downflow respectively.  EIT
also recorded one image using 171~{\AA}, 304~{\AA} and 284~{\AA}
channels in the very late phase of the downflow (see
Fig.~\ref{eit_four}). These images reveal a multi-temperature
behaviour of the downflowing plasma. Since the EIT provided only one
image during the downflow phase using these channels, we could not study the
evolution of the downflow plasma in other channels.

Fig.~\ref{halpha} displays a sequence of images taken by the
H${\alpha}$ coronagraph. The first image was taken at 17:09
UT. Although the ACOS instruments recorded the very early phase (start
phase) of the downflow, they unfortunately missed the eruption
phase. The FOV of the H${\alpha}$ observation is ideal for an
investigation of the origin and evolution of this downflow. Although,
the H${\alpha}$ coronagraph provides data with a regular cadence of
about three minutes, we only show some selected images in
Fig.~\ref{halpha}. For the complete sequence in animation format see
movie 'halpha.mov'\footnote{The movies are available online.}. At
17:09~UT (top left frame in Fig.~\ref{halpha}) a lot of material is
piled high up in the corona with some bright threads still connected
to the Sun's surface. As time passes, some of the material seems to
move away and might have escaped along with the CME. In addition some
material moves downward.

After a while (at 17:48~UT) multiple bright localized structures
(cusp-shaped features) can be seen at the location (950, 1200; in
arcsec). These localized features are marked as 'S1', 'S2' and
'S3'. The feature on the right, namely 'S1', is brightest. These three
features exist until 17:54~UT. After that only one, very bright
(brighter than the earlier three) cusp-shaped feature remains, which
then propagates downward. The two branches in the downflow, marked
'B1' and 'B3', emanate from this cusp and material flows down along
these two branches before the right branch 'B1' bifurcates into
another branch namely 'B2' at 18:07~UT. Interestingly this location
and time corresponds to the kink in the right branch seen in the EUV
images (see Fig.~\ref{eit_downflow}). The two branches 'B1' and 'B3'
of the downflow are also evident in the EUV observations, though the
left 'B3' branch is not as clearly discernible as in H${\alpha}$
observations. The third branch of the downflow 'B2', which is bright
and strong in H${\alpha}$ images, is not evident in the EUV images,
most likely because the plasma flowing along the third branch does not
radiate in the narrow temperature range to which the EIT 195~{\AA} channel
is sensitive. The left branch ('B3') had almost disappeared by
18:37~UT and the middle branch ('B2') disappears at 18:28~UT. However,
the kink in the right branch - where the middle branch emanates -
remains clearly visible. Most of the material seems to flow down along
the right branch of the downflow feature, which is the brightest and
longest lasting.
\begin{figure*}
\centering
\includegraphics[width=0.8\textwidth]{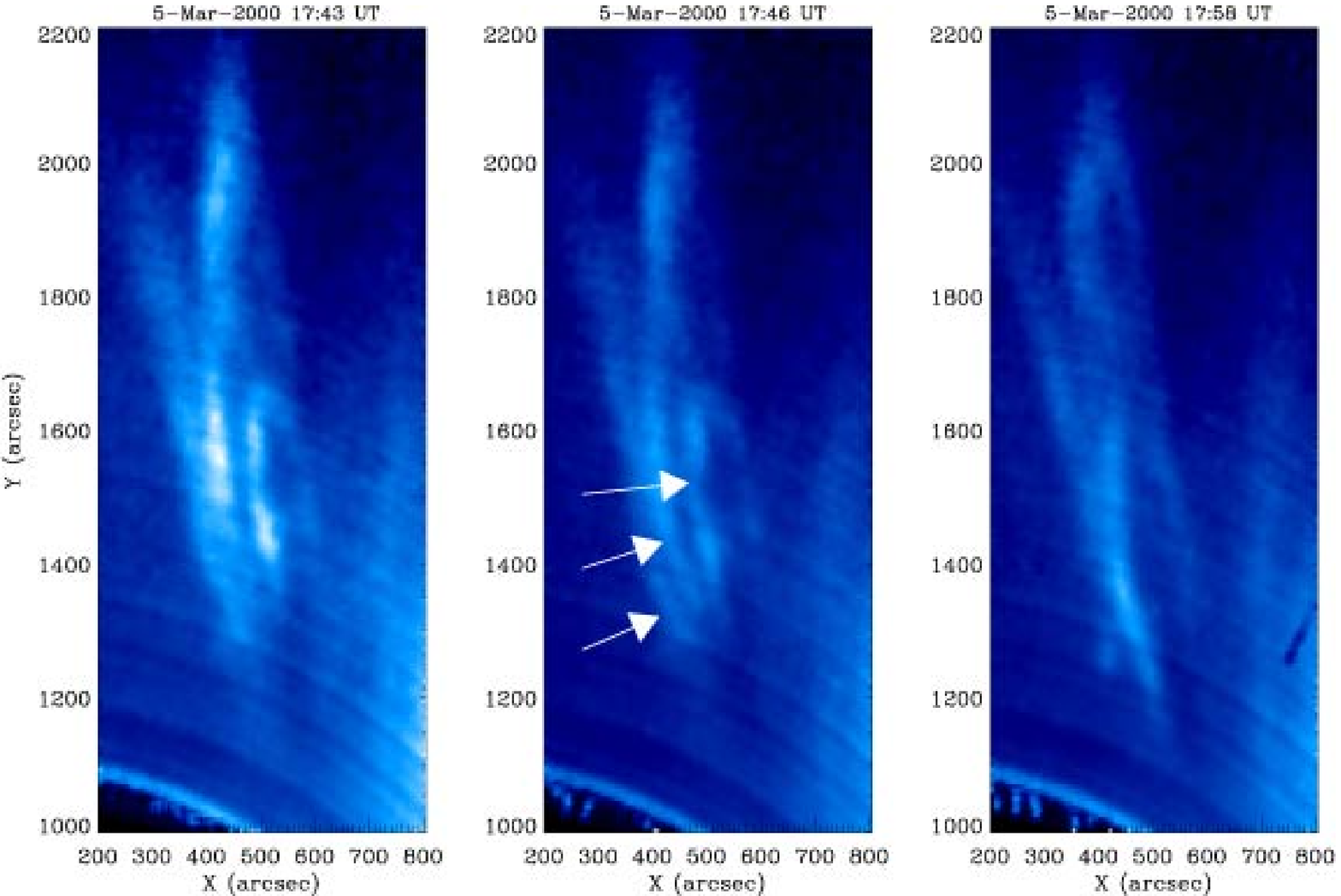}
\caption{Absolute intensity images taken by Mk4 white-light
coronagraph. The arrows in the middle panel mark the shrinking
loops.\label{abs_k4}}
\end{figure*}
\begin{figure}
\centering
\includegraphics[width=0.6\textwidth]{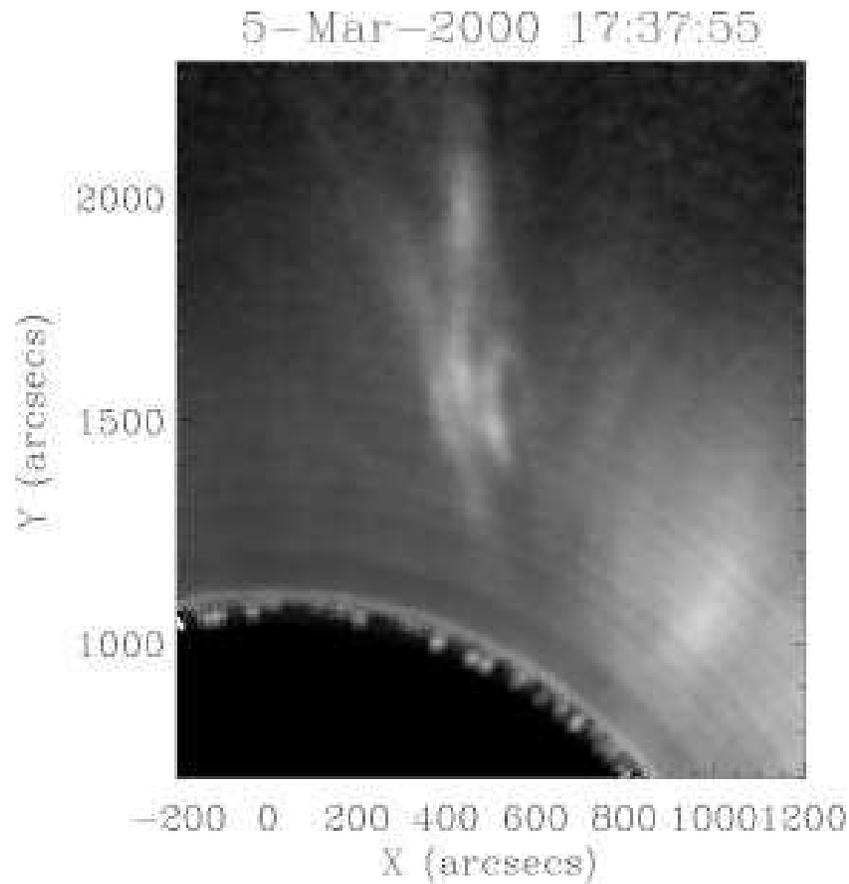}
\caption{Images recorded by the Mk4 which was used as an base image in
creating Fig.~\ref{Mk4}.\label{base}}
\end{figure}
\begin{figure*}
\centering
\includegraphics[width=0.8\textwidth]{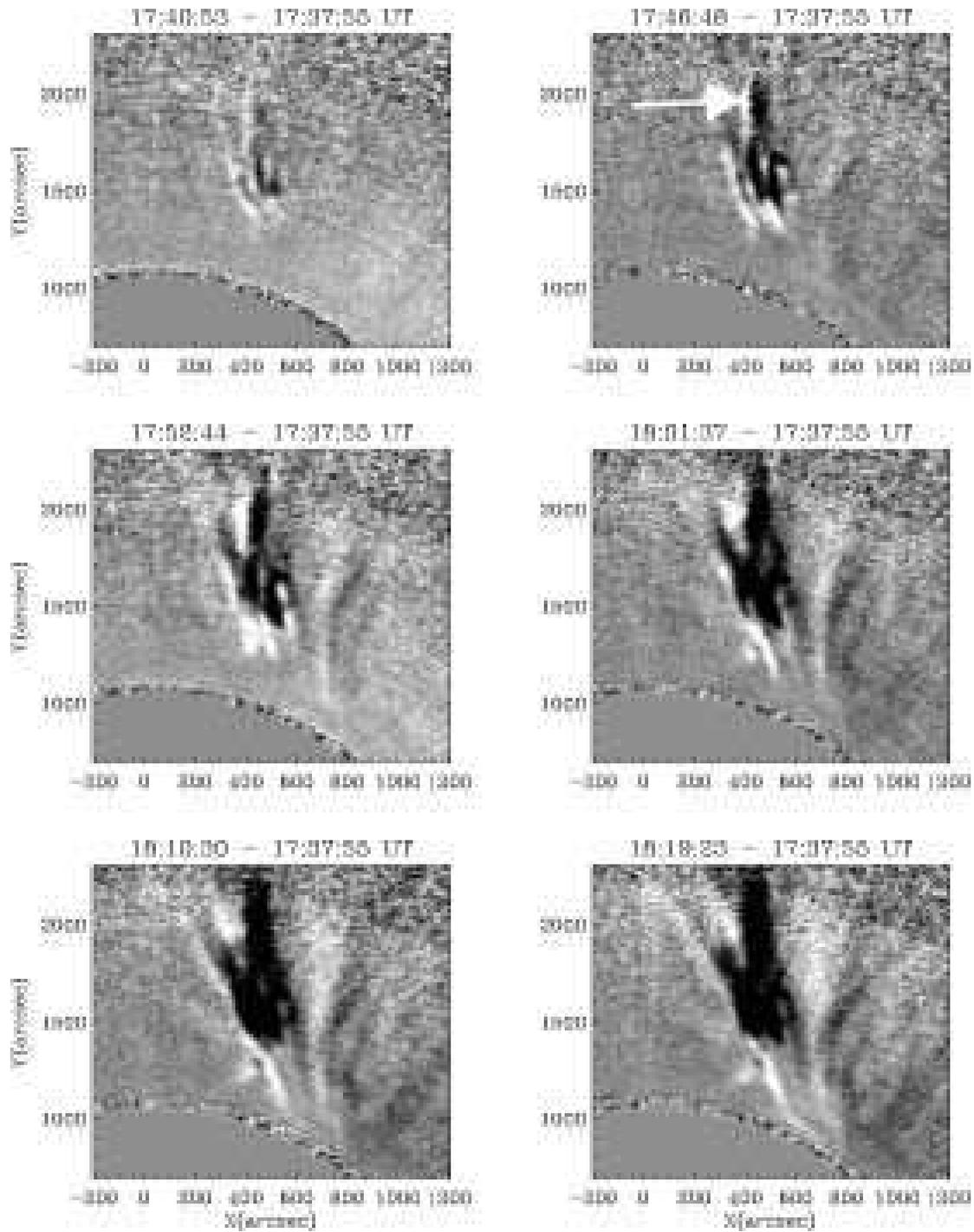}
\caption{Base difference images taken by the Mk4 coronameter of the
Advanced Coronal Observing System at Mauna Loa Solar Observatory. The
arrow in the top right panel locates the outward propagating
reconnetion jet. Note that only few images are shown here. For the
complete sequence see the movie 'mk4.mov' (online only).\label{Mk4}}
\end{figure*}
\begin{figure}
\centering
\includegraphics[width=0.45\textwidth]{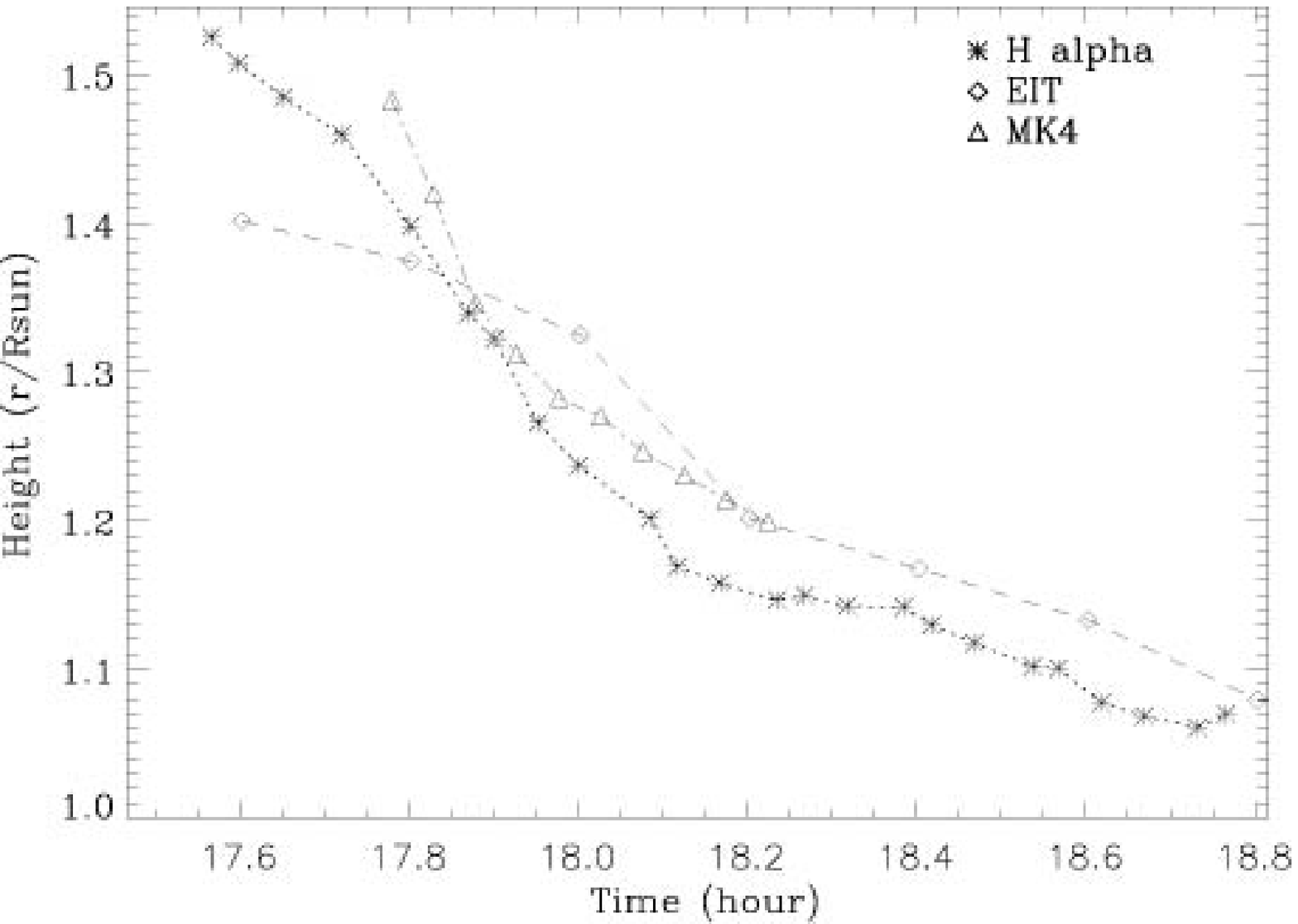}
\caption{Height-time plot of the right branch (B1) of the downflow
based on observations recorded by the $H{\alpha}$ coronagraph
(asterisks), Mk4 (triangles) and EIT 195~{\AA} (diamonds). Note that
the data points for Mk4 were obtained from base-difference images. EIT
data points are taken from Tripathi et al. (2006).\label{ht_all}}
\end{figure}

In order to compare our EIT observations with those made in
H${\alpha}$, we display in Fig.~\ref{eit_halpha} three EIT images
co-aligned with H${\alpha}$ images recorded very close in time (top
panel: EIT-17:10:48~UT, H${\alpha}$-17:09:58~UT; middle panel:
EIT-17:58:50~UT, H${\alpha}$-17:57:09~UT; bottom panel:
EIT-18:23:03~UT, H${\alpha}$-18:23:06~UT). Since the H${\alpha}$
coronagraph does not provide full disk observations like EIT, it is
not straightforward to co-align these images. For the co-alignment, we
co-registered the EIT images to the near-simultaneous H${\alpha}$
images using the routine {\it coreg\_map.pro} provided in the SSW
tree\footnote{http://ydac.mssl.ucl.ac.uk/sswdoc/solarsoft/}. This
routine is a wrapper around another routine called {\it drot\_map.pro}
which differentially rotates one map at the time of the other map,
while taking into account the roll angle. Also, in order to have same
pixel size in two maps, images with a smaller pixel size are rebinned
to a high number of pixels. The co-registration provides the EIT
images with the same pixel size (2.9 arcsec) as that of H${\alpha}$
images. The H${\alpha}$ contours representing the right branch of the
downflow (right panel of Fig.~\ref{eit_halpha}) were seen to be
spatially and temporally correlated with those of EIT. In the left
branch the H${\alpha}$ brightenings appear to be far more localized
and point-like than the more thread-like structures seen in EIT. These
bright H${\alpha}$ points do, however, correspond to EIT
brightenings. It seems plausible and reasonable to conclude that the
features observed in EIT and those in H${\alpha}$ are closely
connected, although due to different temperature sensitivity of the 2
data sets it is likely that the plasma is composed of multi-thermal
unresolved magnetic strands. The cusp-shaped feature seen in
H${\alpha}$ is located outside the FOV of EIT. Moreover, downflow
branches in the EIT and the apex location where the two branches
emanate in EUV images seem to be wider than that in H${\alpha}$.

The top panel of Fig.~\ref{mask} displays an H${\alpha}$ map with a
masked region. We selected this region in order to compute the
variation of the total amount of material during the sequence of the
downflow which is shown in the bottom panel. It is evident from the
plot (see bottom panel of Fig.~\ref{mask}) that the total intensity of
the masked region increases over most of the time that the downflow
was seen. This could either be due to the increase in the amount of
the downflowing plasma, to an enhancement of the density, or cooling
of hot material to chromospheric temperatures. Further the total
intensity starts to decrease when most of the material has drained down
on to the Sun's surface.

\begin{figure*}
\centering
\includegraphics[width=0.8\textwidth]{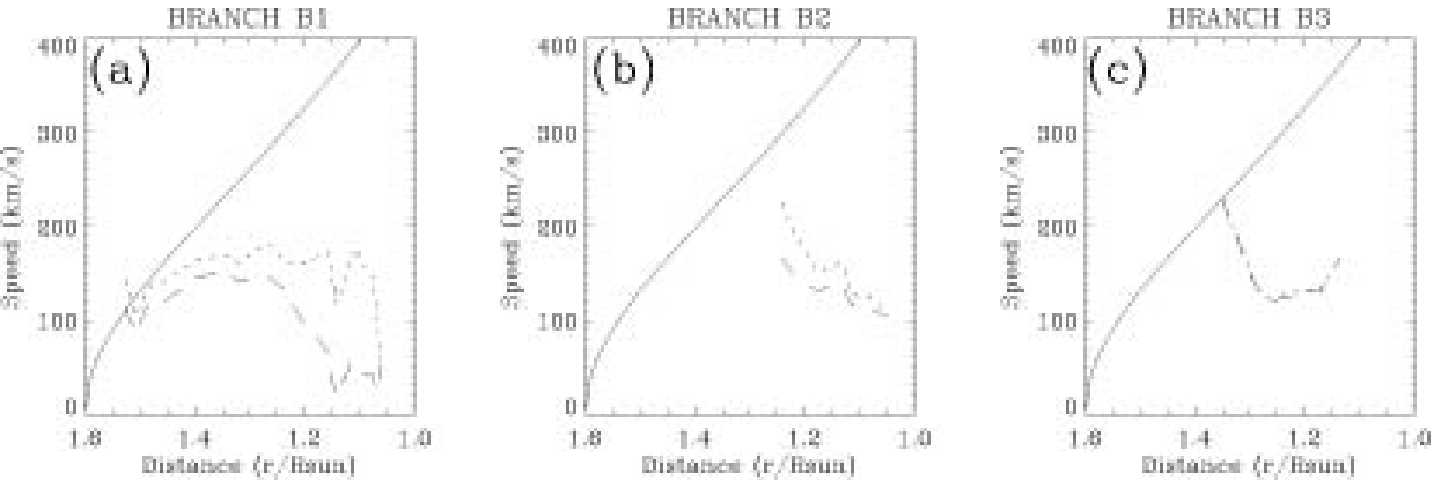}
\caption{Speed-height plots for the downflow (B1, right branch: panel
a; B2, left branch: panel b; B3, left branch: panel c) based on the
$H{\alpha}$ coronagraph observations. Dashed lines represent the
absolute projected speed and dotted lines represent the downward
component of the speed. The solid lines represent the free-fall speed
profile of a body starting at 1.6 R$_{\sun}$. The variation of
gravitational acceleration with height was taken into
account.\label{speed_halpha}}
\end{figure*}

The downflow was also recorded in white-light observations taken by
the Mk4 coronameter of the ACOS at MLSO. The advantage of Mk4 is the
larger FOV. Combining the EIT, the H${\alpha}$ and the Mk4
observations provides an opportunity to study the solar corona out to
2.79 R$_{\sun}$. Since white-light images can be directly interpreted
in terms of the distribution of electron density along the line of
sight, they potentially contain information on the dynamics of the
magnetic field.  Fig.~\ref{abs_k4} displays absolute intensity images
taken by the Mk4 coronagraph. Three bright streak like structures are
evident and are marked in the middle panel with arrows. These
structures appear quite discrete in the left panel. With time, these
streaks diffuse and most probably move down.  Fig.~\ref{Mk4} displays
the sequence of base difference images taken by the Mk4
coronameter. Base difference images provide information about the
dynamics of features with respect to a fixed image frame. In this case
the image recorded at 17:37:55~UT (see Fig.~\ref{base}) was taken as
the base image. We chose this particular image as a base since the
leading edge of the downflow started to be clearly discernible at this
time. Although Mk4 provides observations with a cadence of about 3
minutes, we only show some selected images in this figure. For the
complete sequence see the movie 'mk4.mov'\footnote{Movies are
available online}. Three distinct features can be seen in the top left
images in Fig.~\ref{Mk4}. As time passes it appears that the complete
structure is breaking into two parts. One moving outwards (see bright
feature at located 500, 1900 arcsec location in the top right image
taken at 17:46~UT) and the other moving downwards (see bright feature
at 500, 1400 arcsec in the top right image). The downward moving
feature appears initially to be blob-like and starts to bifurcate at
17:52:44~UT (more clearly at 17:55:41~UT and afterwards) at a point
which corresponds to branches 'B1' and 'B3' seen in H${\alpha}$ (see
Fig.~\ref{halpha}). The branching of the downflow was apparent in
H${\alpha}$ at 17:54:04~UT becoming more clear at 17:57:09~UT. The
bifurcation of the right branch could not be recorded by Mk4 because
of the occulting disk. The large dark area, or dimming, that slowly
builds up in the center of the image during the sequence is due to a
general decrease in the brightness there as the body of the CME and
embedded prominence moves.  As can be deduced from Fig.~\ref{Mk4} and
more clearly from the movie 'mk4.mov', the dimming area increases in
all four directions, but predominantly in the direction of the outflow
and downflow. The increase in the dimming area can be interpreted in
terms of reconnection as proposed by \cite{siota} based on MHD
modelling.
\begin{figure*}
\centering
\includegraphics[width=0.60\textwidth]{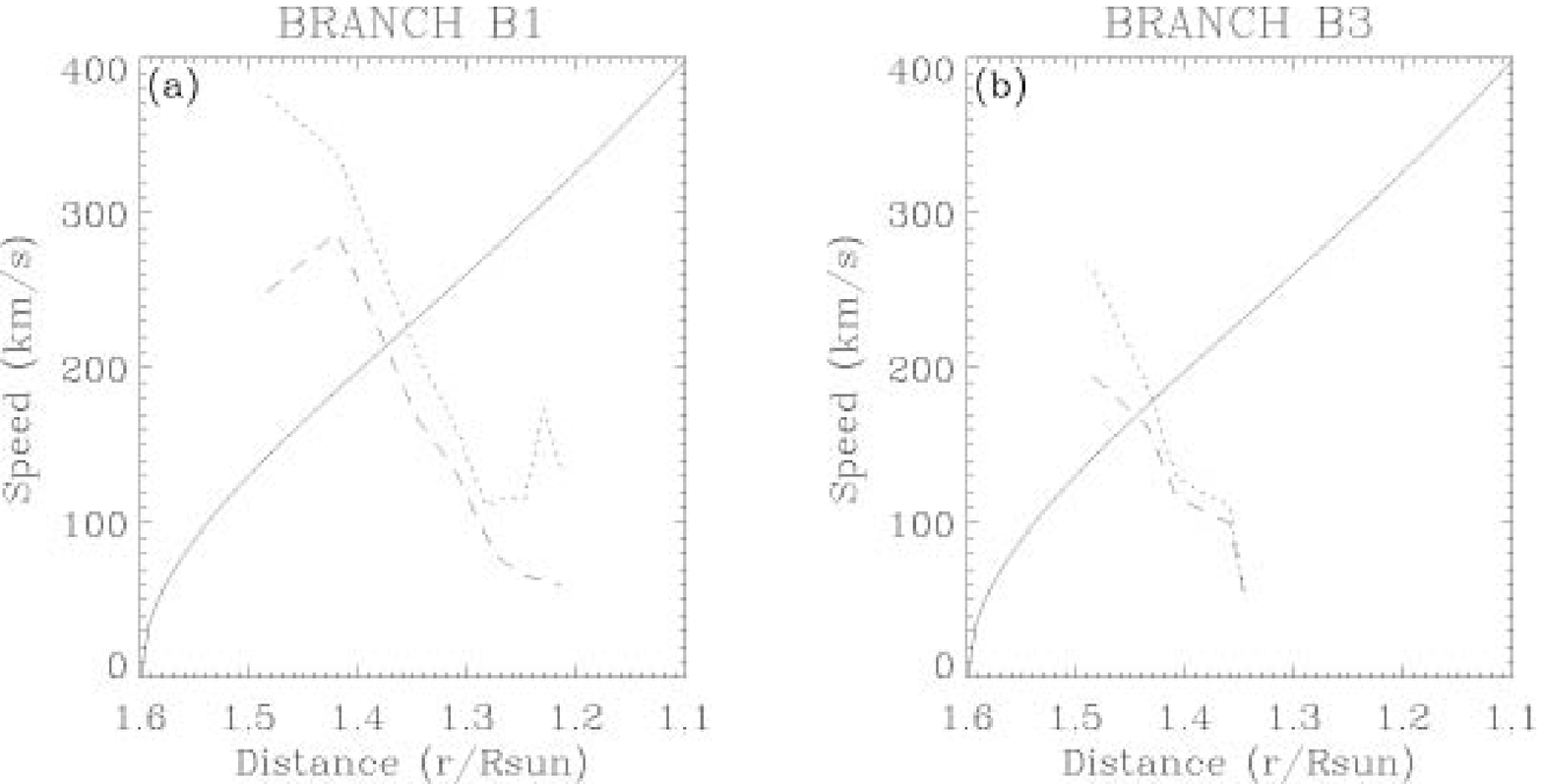}
\caption{Speed-height plots for the downflow (B1, right branch: panel
a; B3, left branch: panel b) based on the Mk4 coronagraph
observations. Dotted lines represent the absolute projected speed and
dashed lines represent the downward component of the speed. The solid
lines represent the free-fall speed profile of a ballistic body
starting at 1.6 $R_{\odot}$. The variation of gravitational
acceleration with height was taken into account. \label{speed_k4}}
\end{figure*}
\section{Height-time and speed-height measurements}

Figure~\ref{ht_all} displays the height-time plot for the right branch
'B1' of the downflow as measured from the observations recorded by the
H${\alpha}$ coronagraph (asterisks), Mk4 coronameter (triangles) and
EIT 195~{\AA} (diamonds). Note that the data points from Mk4 are
obtained from the base difference images (since in the original
images the features are rather diffuse and it is easier to follow the
dynamics of features using base difference images) and those for EIT
are taken from Paper I. These data points refer to the leading edge of
the features. We did not attempt to compare the height-time diagram
for other branches as they were not observed with all instruments
simultaneously. The height-time diagram for the right branch matches
very well for all three instruments except the very first point of
EIT. This could be due to an error in the position of obtaining the
first EIT data point as it was quite weak and lay right at the edge of
the image frame. In general the gas falls more rapidly in the early
phases and more slowly in the later phases.

Figure~\ref{speed_halpha} displays the speed-height profiles of the
right branch (B1, panel a), middle branch (B2, panel b) and left
branch (B3, panel c) as obtained from the H${\alpha}$
observations. The dashed line represents the downward component of the
speed and the dotted line indicates the absolute projected speed for
the downflows. In order to avoid excessive fluctuations in the speed
we have applied a smoothing (running mean) to the data points as we
are mainly interested in the general trend of speeds rather than their
local fluctuations. Note that all the measured speeds represent a
lower limit on the real speed, since the direction of motion may lie
outside the plane of the sky. The solid lines represent the
speed-height profile of a ballistic body falling from a height of 1.6
R$_{\odot}$, taking only solar gravitation into account. The free fall
height was chosen to be 1.6 R$_{\sun}$ as the movie 'mk4.mov' reveals
that the bifurcation would have happened at around this height.  It is
evident from Fig.~\ref{speed_halpha} that the right and the left
branches of the downflow seen in H${\alpha}$ (panels a \& c) started
with almost the free-fall speed and were strongly
decelerated. However, the middle branch downflow (panel b) had a speed
well below free fall. This is clearly due to the fact that this branch
bifurcated from the right branch at a time when the latter was already
highly decelerated.

Fig.~\ref{speed_k4} displays the speed-height profile for the right
branch (B1, panel a) and that of the left branch (B3, panel b) from
Mk4. The different lines have the same meaning as in
Fig.~\ref{speed_halpha}. Recall that the data points for Mk4 were
obtained from the base difference images. There are no measurements
for the middle branch as it was below the occulting disk of Mk4. The
initial speeds measured from the base difference Mk4 images are much
higher than those obtained from the H${\alpha}$ absolute intensity
images. The downflow started with a much higher speed ($\approx$ 380
km~s$^{-1}$) than free fall, but then rapidly decelerated to values
well below free-fall. These later speeds are in reasonable agreement
with those found from H${\alpha}$. The right and left branches show
similar speed profiles, although the speed of the left branch downflow
is lower than that of the right branch.

Based on the MHD treatment of magnetic reconnection
(i.e. Sweet-Parker and Petschek type reconnection), the reconnection
outflow speed appears to be roughly equivalent to Alfv\'en speed. If
the total magnetic energy stored at the reconnection location converts
into the bulk kinetic energy of the plasma, the outflow speed become
equal to the Alfv\'en speed \citep{yokoyama}. Considering the magnetic
field in prominences to be about 20 G \citep[see e.g.,][]{casini} and
the density as 10$^{10}$~cm$^{-3}$ \citep[see e.g.,][]{chang}, the
Alfv\'en speed is then about $\approx$440~km~s$^{-1}$. This is larger
than the measured initial speed of the downflow. Taking into account
that some of the energy converts into the thermal energy heating the
plasma, the initial speed of the downflow appears to be consistent
with a reconnection outflow.

\section{Summary and discussion}

We have investigated the multi-wavelength observation taken by
SoHO/EIT at 195~{\AA}, and the ACOS instruments namely, the PICS H${\alpha}$
coronagraph and the Mk4 white-light K-coronameter in order to study the
origin, evolution and characteristics of a {\sl bright} coronal
downflow. The downflow was observed after three prominences
simultaneously erupted as part of a CME on 5~March 2000. Although
there were three prominence eruptions (two large (P1 and P2) and one
small (P3); see Fig.1 in Paper~I), only one CME was detected by
LASCO/C2 coronagraph. The two large prominences (P1 and P2) were
identified in LASCO/C2 images as the core of the CME (see Fig.2 in
Paper I).

\subsection{Summary of results}

\begin{itemize}

\item The downflow is seen in the EIT (195~{\AA}, 171~{\AA},
 284~{\AA}, 304~{\AA}), SXT (see Paper I), ACOS H${\alpha}$
 coronagraph and Mk4 white-light coronameter observations, so that it
 must comprise multi-thermal plasma.

\item The downflows observed by different instruments appears to be
approximately co-spatial and co-temporal, although the cool
H${\alpha}$ gas lies somewhat below the hotter EIT gas most of the
time (see Fig.~\ref{ht_all}). The larger FOV of the Mk4 coronameter
relative to that of the EIT allowed us to determine that the downflow
started roughly at 1.6 R${\sun}$.

\item The Mk4 base difference images show a blob-like feature pushed
downward at around 17:46~UT which later bifurcated into two branches
namely 'B1' and 'B3' (see Fig.~\ref{Mk4}, top right
panel). Simultaneously, a bright outward moving feature was seen (see
Fig.~\ref{Mk4}, images in the second and third row). For a complete
sequence see movie 'mk4.mov' (available online).

\item The kink in the right branch (branch 'B1') of the downflow was
seen at 18:12~UT in EIT observations and at 18:05~UT in H${\alpha}$
coronagraph observation. The middle branch (branch 'B2') of the
downflow emanated at the location of the kink. This branch is only
seen in H${\alpha}$ observations. In the EIT, the middle branch was not
clearly observed, although an bright streak emanating from the kink
location is apparent in the running difference images shown in the
third image in the first column of Fig.~\ref{eit_downflow}.

\item The H${\alpha}$ images show many bright streaks which are
probably aligned along the magnetic field. At least three can be seen
clearly and are labelled 'S1', 'S2' and 'S3' in the image shown in the
second row first column in Fig.~\ref{halpha}. The Mk4 white light
images show a bright shrinking loop-like structures with cusp-shaped
loop top (last 3 frames of Fig.~\ref{Mk4}).

\item The H${\alpha}$ observations also reveal that the amount of the
downflowing plasma at low temperature increases over most of the time
that the downflow was seen. This is suggestive of condensation of
plasma from high coronal temperatures. Further the amount of
H${\alpha}$ plasma starts to decrease when most of the material drains
down to the Sun's surface.

\item The height-time plot (see Fig.~\ref{ht_all}) of the right branch
of the downflow shows a similar profile for all three different
instruments, with hot plasma (see EIT) lying above the cooler
plasma. The height-time profile clearly reveals the kink in space and
time. Note that we only produced the height-time plot for the right
branch as this was observed with all three instruments simultaneously.

\item Speed-height diagrams reveal that the downflow seen in
H${\alpha}$ started with almost free-fall speed, but then slowed
(Fig.~\ref{speed_halpha}). However, Mk4 observations
(Fig.~\ref{speed_k4}) show that the initial downflow speed was much
higher than the free-fall speed. We note that Mk4 (white-light
coronagraph) is sensitive to plasma at all temperatures, so the hot
plasma might be moving faster than the cool plasma.

\end{itemize}
\subsection{Discussion}

Let us consider possible explanations for the origin of the downflow
and its characteristics. One possibility is that the downflow is
composed of material that could not reach the escape speed and
overcome the solar gravitational field. Under this scenario the speed
of the downflowing plasma should not exceed that of free fall. It can
be lower, since lower lying, possibly still upward moving, plasma
would slow down the downflowing plasma. The speed obtained from the
H${\alpha}$ observations is comparable to that of free-fall or lies
below it. However, the initial downflow speed obtained from the Mk4
recording lies far in excess of the free-fall speed. Note that the
speeds are measured in the plane of the sky so that they are lower
limits. Furthermore, the speeds derived from the Mk4 observations have
the advantage over those from the H${\alpha}$ or the EIT observations
because the Mk4 observations are not temperature sensitive. Therefore
the derived data points based on the Mk4 observations are not
temperature biased. There may, however, be an enhanced uncertainty due
to the fact that we are analyzing base difference images of Mk4
data. We note that, the downflow was observed in EIT (1.5~MK) and SXT
(2-5~MK) images (see Paper I). This is suggestive of plasma heating
which is highly unlikely if the downflow is simply due to
deceleration/acceleration by the solar gravitational field.

Another interpretation is that while erupting the prominences pass
through a kink instability. The development of a kink in prominences
during the eruption phase has been observed
\citep[e.g.,][]{david_kink, rust_kink} and theoretically modelled
\citep[e.g.,][]{tibor_kink, fan_kink}. The kink instability in the
prominence during eruption can explain the cusp-shaped structure
formation which is seen in the images taken by the $H{\alpha}$ and
the Mk4 coronagraphs as well as the EIT. Furthermore, during the
eruption the helical field lines, originally holding the prominence at
the bottom, get stretched and the material sitting at the bottom of
the flux rope drains down along the legs of the flux rope. This
interpretation poses similar problems to the earlier one, such as the
plasma heating and the high speed of the downflow with respect to
free-fall.

Let us now consider the interpretation put forward by \citet[based on
the MLSO H${\alpha}$ coronagraph observations]{holly00, holly01} and
later \citet[based on 3D MHD simulations]{sarah_part07, sarah_part06}.
These authors proposed that during the eruption of a pre-existing flux
rope, reconnection takes place internally and the flux rope breaks
into two. The outer part propagates along with the CME and the lower
part falls back to the Sun's surface. The basic observables predicted
on the basis of the above model would be the simultaneous presence of
an X-type structure in the corona, a three part structured CME and the
downflowing plasma \citep{trip_sarah}. One of the most important and
plausible signatures of the reconnection would be heating of plasma
emanating from the reconnection locations (X-type location) as well as
reconnection jets. This implies that the speed of plasma emanating
from the reconnection location and flowing towards the Sun's surface
would be significantly higher than the free fall speed especially at
high temperature. This speed would either grow or decrease depending
on the amount of material below the location of reconnection.

Based on the LASCO/C2 and C3 observations we confirmed that the
associated CME was comprised of a bright front, dark cavity and a
bright core - representing the prominence material (see Paper
I). Also, the downflowing plasma in the EIT and the SXT was bright,
implying that the temperature of at least a part of the downflowing
plasma can be as high as 4-5~MK or even 20~MK \citep{trip_trace},
providing strong evidence of plasma heating (see Paper I). However,
due to the restricted FOV of the EIT and the SXT and lack of
observations from 1.5-2.5 R$_{\sun}$, we were not able to locate the
precise reconnection point, reconnection jet and estimate the speed of
the downflowing plasma in the early phase. The Mk4 and the H${\alpha}$
data helped us to find the location of reconnection because of the
larger FOV of these instruments. Based on the Mk4 observations we have
a more complete picture of the reconnection such that the reconnection
jet propagates outwards (see movie 'mk4.mov') and the initial
speed of the downflowing plasma is substantially higher than that of
free-fall (see Figs.~\ref{Mk4},~\ref{speed_k4}). Furthermore, the
white-light base difference images show that the dimming area
increases at the location where the downflow seems to start. The
increase in the dimming area provides further strong evidence of
reconnection as described by \cite{siota}.

Although this interpretation explains most of the characteristics of
the observed downflow, it still poses a problem concerning the reason
for approximately co-spatial and co-temporal observation of
multi-thermal plasma such as in the H${\alpha}$ (cool material) and
EIT (all four channels) and SXT (hot material). There could be
different possibilities. First, this could be due to the fact that
cooler plasma (lower part of the prominence) seen in the H$\alpha$ is
still rising when reconnection occurs further up. Because of the
reconnection, the plasma is stopped and slowly starts to fall down. In
this scenario material would not be heated to the temperatures needed
to make it visible in the EIT and the SXT. Another possible
explanation is that the EIT and the H${\alpha}$ observations indicate
a dense flux tube-like structure. The highly dense plasma in the inner
part of the flux tube is either cold to start with, or cools down
faster than the outer part, as the radiative cooling time is
proportional to n${_e}$$^2$. This interpretation also explains the
fact the H${\alpha}$ downflow in the left branch is highly localised
and seen at locations where the EIT downflow is brightest. Moreover,
the H${\alpha}$ downflows are thinner than those in EIT. On the other
hand, it may well be that the downflow is comprised of multi-thermal
unresolved magnetic strands. The above two interpretations are also
supported by the fact that the total intensity of the masked region
increases over most of the time when the downflow is seen (see
Fig.~\ref{mask}). This increase is basically due to the enhancement of
the downflowing material, which means that the plasma that was at
higher temperature cools down to temperatures sensitive to the
H${\alpha}$ observations. Later on the total intensity starts to
decrease when most of the cool material is drained down on to the
Sun's surface and there are no hot material left to cool down.

Despite the fact that we can explain the origin and characteristics of
the observed downflow described in this paper and Paper I, the
question remains as to why such downflows are rare event?. We also
looked at five more H${\alpha}$ observations from MLSO.  Draining of
the plasma along the legs of erupting prominences was a common
phenomenon in the H${\alpha}$ observations for all those five
events. These downflows were also seen in the Mk4
observations. However, there were no corresponding signatures in the
EIT observations. Also there was no evidence of formation of an X-type
structure such as a cusp in the corona. This suggest that reconnection
associated with the heating of the prominence gas to coronal
temperatures is not an entirely common occurrence.

We note that this is the first observation of its kind and demands
further study and a deeper understanding. In order to carry out this
kind of study we would require a wide temperature coverage with very
high time resolution data over a large field of view. In the future,
observations from the Hinode satellite combined with those from the
Solar Terrestrial Relation Observatory (STEREO) and later the Solar
Dynamics observatory (SDO) may provide a unique opportunity to study
these phenomena in more detail.

\begin{acknowledgement}
{We acknowledge an anonymous referee for comments which certainly
improved the quality of the manuscript. DT and HEM acknowledge support
from PPARC. DW was supported by Air Force Research Lab Contract
FA8718-06-C-0015. DT would like to thank Sarah Gibson for many useful
discussions and comments. We thank the SoHO-EIT and HAO teams for
providing the data and also Joan Burkepile for her help in explaining
the instruments. SoHO is a mission of international collaboration
between ESA and NASA.}

\end{acknowledgement}

\bibliographystyle{aa}
\bibliography{references}


\end{document}